\documentclass[aps,prl,twocolumn,notitlepage]{revtex4-1}

\usepackage{amstext,amsmath,amssymb,amsfonts,bbm}
\usepackage[latin1]{inputenc}
\usepackage{epsfig}
\usepackage{hyperref}
\usepackage{amsthm}
\usepackage{subfigure}
\usepackage{color}
\usepackage{multirow}
\newcommand{\bea}{\begin{eqnarray}}	
\newcommand{\eea}{\end{eqnarray}}
\newcommand{\cG}{{\cal G}}

\begin{document}

\title{\Large \bf Reply to comment on ``Lost in translation: topological
singularities in group field theory''}

\author{{\bf Razvan Gurau}}\email{rgurau@perimeterinstitute.ca}
\affiliation{Perimeter Institute for Theoretical Physics, 31 Caroline St. N, ON N2L 2Y5, Waterloo, Canada}

\begin{abstract}
    In \cite{Smerlak:2011rd} the author disputes the conclusion of our paper 
\cite{Gurau:2010nd}. He claims that the Feynman graphs of three dimensional 
group field theory always represent pseudo manifolds. 
   However 
   \begin{itemize}
   \item  \cite{Smerlak:2011rd} uses a different definition for pseudo manifolds. 
   \item In order to apply the new definition \cite{Smerlak:2011rd} proposes a construction
   which cannot be implemented in a path integral by Feynman rules.
  \end{itemize}
  These two points invalidate the claims of \cite{Smerlak:2011rd}.
\end{abstract}

\maketitle

Group field theory (GFT) is a quantum field theory defined by some action. Its partition function (and correlations) 
are sums over graphs built in accordance to the Feynman rules. All the information on the graphs is strictly
combinatorial. It is the physicists task to recast the combinatorial data into further structures associated 
to the Feynman graphs. 

The question ``is a graph a pseudo manifold?'' is addressed in both \cite{Gurau:2010nd} and \cite{Smerlak:2011rd}.
Although the author fails to mention it, the definition of pseudo manifold used in 
\cite{Smerlak:2011rd} (paragraph 6: ''[...] a pseudo-manifold is {\bf a topological space}
having [...] '') is {\it different} from the definition used in \cite{Gurau:2010nd} 
(page 6: ``
An n-dimensional simplicial pseudo manifold is {\bf a finite abstract simplicial
complex} [...]
'') (our emphasis).

Are the Feynman graphs of arbitrary group field theories always dual to finite abstract simplicial complexes 
(see again \cite{Gurau:2010nd} for the precise definition)? No, they are not (as the author of \cite{Smerlak:2011rd} also 
mentions it in paragraph 3). Hence they {\bf cannot be} pseudo 
manifolds using the definition of \cite{Gurau:2010nd}. On the contrary, the graphs of the colored models  
\cite{Gurau:2010nd} are (and are in fact dual to simplicial pseudo manifolds \cite{Gurau:2010nd}).

Can one apply the definition of pseudo manifold proposed in \cite{Smerlak:2011rd}? In order to do this one must first 
associate a topological space to a graph. In paragraph 1, the author of \cite{Smerlak:2011rd} reinterprets
the GFT vertices as tetrahedra and the GFT lines as simplicial gluing maps and reconsiders the graph 
as the ``quotient'' of the tetrahedra along the simplicial maps. The problem with this construction lies in 
paragraph 3 of \cite{Smerlak:2011rd} ``[...] This is because the identification of triangles can result
in the folding of edges onto themselves, thereby spoiling the injectivity of the projections on the interior of the
edges.''. What is the quotient topology one builds in this case? As non injective (i.e. not one to one) maps are 
not invertible, in the absence of extra suppositions on these ``foldings'', one cannot claim that the resulting 
space is homeomorphic with anything. 

The only information given by the GFT is that the two end vertices of an edge are identified. 
There is no way (or at least none presented in \cite{Smerlak:2011rd}) one can implement any additional 
information on these foldings in the GFT action (i.e. translate it in Feynman rules). Any further assumption 
one makes at this step cannot come from group field theory. There are at least two prescriptions one can
consider 
\begin{itemize}
 \item The entire edge is identified with its end points. Indeed, the only linear map whose end points are identical 
       is the constant map.
 \item One choses a point belonging to the interior of the edge, divides it in two segments, which one then identifies.
\end{itemize}
 One could also propose other prescriptions, for instance one can chose three points on the edge and identify 
the four resulting segments, etc..

The author of \cite{Smerlak:2011rd} does not address this point. 
The sequel (paragraph 4 ``[...] a sufficient number of barycentric subdivisions to the
tetrahedra prior to their gluing to make sure that none of the new edges get folded onto themselves, 
thus inducing a triangulation on the quotient [13, 15]'') works if one choses the second 
prescription (and in this case leads to a pseudo manifold as defined in \cite{Smerlak:2011rd}),
but {\bf fails} if one choses the first prescription.

Of course one is free to add any number of suppositions to the combinatorial data given by the GFT. For instance one can 
always decide to associate a sphere to all graphs (this cannot be implemented either by Feynman rules, and is obviously 
somewhat arbitrary). The price to pay is that the resulting space is unrelated to the initial graph. Applying 
the second prescription for the graph $\cG^1$, figure 7 of \cite{Gurau:2010nd} for instance, leads to a pseudo manifold 
possessing two vertices with $\mathbb{R}P^2$ links ( see \cite{DePietri:2000ii}, page 17 ``[...]
and it is easy to see that its link is the projective plane [...]''). Such a pseudo manifold is not 
homotopically trivial whereas the graph $\cG^1$ is. Indeed, for this example, is seems safer to always associate a 
sphere to a graph...

We conclude that, as \cite{Smerlak:2011rd} does not use the same definition of pseudo manifold as \cite{Gurau:2010nd}, and
presents a construction which cannot be implemented by Feynman rules, it is not pertinent to the subject
treated in \cite{Gurau:2010nd}.

\end{document}